\documentclass[10pt,journal]{IEEEtran}

\usepackage{textcomp}

\usepackage{amsmath}
\usepackage{amsfonts}
\usepackage{amssymb}
\usepackage{mathrsfs}
\usepackage{color}
\usepackage{hyperref}
\usepackage{booktabs}
\usepackage{graphicx}
\usepackage{epstopdf}
\usepackage{epsfig}
\usepackage{cite}
\usepackage{enumerate}
\usepackage{amsopn}
\usepackage{bm}
\usepackage{algorithm}
\usepackage{algpseudocode}

\newtheorem{lemma}{Lemma}
\newtheorem{theorem}{Theorem}
\newtheorem{remark}{Remark}

\newtheorem{assumption}{Assumption}
\newtheorem{problem}{Problem}

\DeclareMathOperator{\Tr}{Tr}
\DeclareMathOperator*{\argmin}{arg\,min}

\begin{document}

\title{Multi-Party Dynamic State Estimation that Preserves Data and Model Privacy}

\author{Yuqing~Ni,~Junfeng~Wu, \IEEEmembership{Senior~Member,~IEEE},~Li Li, \IEEEmembership{Member,~IEEE},~and~Ling~Shi, \IEEEmembership{Senior~Member,~IEEE}
\thanks{Y. Ni and L. Shi are with the Department of Electronic and Computer Engineering, The Hong Kong University of Science and Technology, Clear Water Bay, Kowloon, Hong Kong (e-mail: {yniac@connect.ust.hk}, {eesling@ust.hk}).}
\thanks{J. Wu is with the College of Control Science and Engineering, Zhejiang University, Hangzhou, China (email: {jfwu@zju.edu.cn}).}
\thanks{L. Li is with the College of Electronics and Information Engineering, Tongji University, Shanghai, China (email: {lili@tongji.edu.cn}).}
}


\maketitle

\begin{abstract}
In this paper we focus on the dynamic state estimation which harnesses a vast amount of sensing data harvested by multiple parties and recognize that in many applications, to improve collaborations between parties, the estimation procedure must be designed with the awareness of protecting participants' data and model privacy, where the latter refers to the privacy of key parameters of observation models. We develop a state estimation paradigm for the scenario where multiple parties with data and model privacy concerns are involved. Multiple parties monitor a physical dynamic process by deploying their own sensor networks and update the state estimate according to the average state estimate of all the parties calculated by a cloud server and security module. The paradigm taps additively \emph{homomorphic encryption} which enables the cloud server and security module to jointly fuse parties' data while preserving the data privacy. Meanwhile, all the parties collaboratively develop a stable (or optimal) fusion rule without divulging sensitive model information. For the proposed filtering paradigm, we analyze the stabilization and the optimality. First, to stabilize the multi-party state estimator while preserving observation model privacy, two stabilization design methods are proposed. For special scenarios, the parties directly design their estimator gains by the matrix norm relaxation. For general scenarios, after transforming the original design problem into a convex semi-definite programming problem, the parties collaboratively derive suitable estimator gains based on the alternating direction method of multipliers (ADMM). Second, an optimal collaborative gain design method with model privacy guarantees is provided, which results in the asymptotic minimum mean square error (MMSE) state estimation. Finally, numerical examples are presented to illustrate our design and theoretical findings. 	
\end{abstract}


\IEEEpeerreviewmaketitle

\section{Introduction}\label{sec:introduction}
\subsection{Background}
\IEEEPARstart{C}{yber-pyhsical} systems (CPSs) have nowadays been seen in numerous applications, including smart power grid, autonomous vehicles, intelligent transportation, healthcare, and environment monitoring, etc.~\cite{kim2012cyber}.
Privacy issues arise when information that is generated by some party needs to be shared with and stored by other parties. For the purpose to have better understanding of dynamic system states and further manipulate its input to make the system output behave in some desired manner, sensors are deployed at the sites of interest to sequentially collect measurements and send them to remote devices.
In some cases, measurements are privacy-sensitive because they directly contain sensitive information. For example, measurements could be medical records of personal physiological data, such as the skin temperature and blood pressure of patients in a smart hospital. In other cases, measurements are private in that one can infer sensitive information from them.
A particular example is that the habits, financial status, and more of a household could be learned from the daily power data recorded by a meter~\cite{rouf2012neighborhood}. If some cybercriminals obtain and then exploit these data in an unlawful manner, it would probably cause serious consequences~\cite{tonyali2015secure}.

The privacy of data has been extensively investigated in various research fields, such as data science~\cite{kim2003multiplicative,evfimievski2004privacy}, machine learning~\cite{zhang2016dynamic,aono2017privacy,wang2019privacy}, control and systems~\cite{jia2017privacy,wang2017differential,farokhi2017secure}, etc., where the encryption acts as a mainstreaming technique to maintain the confidentiality of data. 
Other than the matters of direct data leakage, in the process of utilizing data to extract meaningful insights or make decisions, there also exists the risk of hidden information disclosure, yet having not drawn proper attention it deserves. 
For example, Tram{\`e}r \emph{et al.}~\cite{tramer2016stealing} demonstrated the possibility of simple and successful model extraction attacks against popular machine learning models. 
The hidden privacy disclosure is a common challenge when we control physical systems, perform parallel computation, or apply artificial intelligence algorithms on shared computing platforms, and cannot be fully overcome only by the traditional encryption schemes. 
For instance, the average consensus is a fundamental algorithm for distributed computing and control, where nodes in a network constantly communicate to learn the average of their initial values in an iterative manner. There could be an undesirable disclosure of a node's initial state to other nodes. Mo and Murray~\cite{mo2016privacy} discussed the initial-value disclosure under the maximum likelihood estimation theory where random noises are used to blur the consensus process. Liu \emph{et al.}~\cite{liu2018gossip1,liu2018gossip2} proposed privacy-preserving gossip algorithms with consistent summation of network node values. They showed that even if eavesdroppers possess the full network structure and flow knowledge, they are unable to reconstruct the network initial node values.
\subsection{Our Work and Contributions}
We focus on dynamic state estimation which harnesses a vast amount of sensing data harvested by various parties and takes place on a third-party computational platform, termed \emph{the secure multi-party dynamic state estimation problem}. The computing framework is economically motivated by the scale advantage, theoretically rooted in the distributed dynamic state estimation and data fusion, and technically rooted from cloud computing and Internet-of-Things, with the purpose to protect the privacy of sensitive information. 
Specifically, multiple parties monitor a physical process by deploying their own sensor networks, respectively. These parties may be companies in competition with each other, or institutions whose data are protected by law. The key point is that they may achieve better estimation performance if they can obtain additional information from others. On one hand, they understand that sharing information is good for all, but on the other hand, they are unwilling, or unable to expose their sensitive information to other parties or a third party without trustworthiness. 

The following example of dynamic state estimation in power grid unveils that the concept of privacy preserving should not only be confined to data, but also be extended to key model parameters. In power systems, the technological advances enable sensors to promptly track the dynamics of the system states. The state-transition model is assumed to be linear~\cite{zhao2019power}:
\begin{align}\label{eqn:dynamic}
x\left(k+1\right)=A x\left(k\right)+w\left(k\right),
\end{align}
where $x\left(k\right)$ is the vector of the bus voltage magnitude and angle at time step $k$, $w\left(k\right)$ captures the state noise, and $A$ characterizes the discrete-time dynamic system model. As a consequence of the power industry restructuring, some of the grid have been administered by independent system operators~\cite{wu2005power}. Operators in different areas observe the system state by deploying their own sensing devices at buses, e.g., phasor measurement units and meters. The observation model for operator $i$ is assumed to be linear:
\begin{align}\label{eqn:observation_equation}
y_i(k)=C_i x(k)+v_i(k),
\end{align}
where $C_i$ is the linearized observation matrix and $v_i(k)$ is the measurement noise. These operators may obtain a more accurate real-time estimate of the overall grid if they could share data. However, operators in different areas are reluctant to share network data and measurements due to competition. The internal measurements and internal line parameters of each area are hidden from all other areas~\cite{kashyap2016privacy}. They worry about that the conventional collaborative dynamic state estimation may lead to the leakage of their trade secrets and potential competitors would learn their technical competence. 

The state estimation problem where a group of sensors monitor a dynamic process has been extensively investigated in  centralized~\cite{shi2014event,song2015multi,yang2017multi} and distributed~\cite{ribeiro2006soi,olfati2007distributed,carli2008distributed,cattivelli2010diffusion,battistelli2014consensus} settings. When there are privacy concerns in the context of state estimation, Gonzalez-Serrano \emph{et al.}~\cite{gonzalez2014state} developed an extended Kalman filter based on encrypted measurements, which was combined with additively \emph{homomorphic encryption}. They considered a scenario with a data owner who provides privacy-protected measurements, and an algorithm owner who processes the encrypted data to estimate the process state.
Based on a similar setup, Zamani \emph{et al.}~\cite{zamani2018private} proposed a secure Luenberger observer combined with the Paillier encryption such that the estimation could be performed on the encrypted data directly. 
Song \emph{et al.}~\cite{song2019compressive} utilized compressive privacy schemes to prevent the fusion center from inferring private states accurately while still allowing it to estimate public states with good accuracy.
In summary, the existing literature mainly focused on the data privacy. However, our vision in multi-party state estimation, for example, the aforementioned dynamic state estimation in power grid, is that the sensitive information to be protected consists of not only local data to be shared but also the parameters of observation models, which has been seldom discussed in the literature, to the best of our knowledge. 

In our paper, we develop a paradigm of privacy-aware multi-party dynamic state estimation. The notion of privacy contains data and model privacy.
In particular, each party's sensitive information, including their locally generated estimates and key parameters of the observation models (i.e., $C_i$ and $R_i$), is kept secret.
To the best of our knowledge, it is the first time that model privacy is taken into account in the context of multi-party dynamic state estimation.
In the paradigm we demonstrate how a fusion protocol that taps additively \emph{homomorphic encryption} enables the parties to jointly estimate the dynamic process while preserving their own data privacy. In general, the data fusion is accomplished by a cloud server and security module. The data producers (i.e., the multiple parties) upload encrypted messages to the cloud server, and the cloud server performs arithmetical operations in the ciphertext space and pass the calculated result to a security module. Finally, the security module is responsible for decrypting and sending back the result to all the parties. We also demonstrate how the parties together develop a stable (or asymptotic MMSE) fusion rule without revealing observation model parameters to others. This procedure is essentially a stabilization problem for a linear filtering system which needs to be solved in a collaborative manner with privacy awareness. 

The contribution of this paper is multi-fold. 
\begin{enumerate}
	\item We highlight the importance of model privacy, i.e., protecting key parameters of the observation models in some multi-party state estimation applications, as the observation models may also be privacy-sensitive. The concept of model privacy, which has begun to be recognized in machine learning communities, has been introduced to dynamic state estimation for the first time in this paper. The notation of privacy considered in this work is two-leveled, and it indicates the need of careful state estimation algorithm design for preserving both levels.
	\item We investigate the estimator stabilization of the proposed paradigm. The fundamental stabilization problem of a linear filtering system is resolved in a collaborative manner with model-privacy awareness. Specifically, methods based on the matrix norm relaxation and ADMM are designed for multiple parties to collaboratively decide their estimator gains without disclosing model parameters.
	\item We also propose an optimal gain design method with model-privacy awareness to achieve the asymptotic MMSE estimation. Essentially, it aims at obtaining the optimal steady-state Kalman gain without disclosing model parameters. To achieve this, we formulate a semi-definite programming problem and solve it by ADMM.  
\end{enumerate}

The remainder of this paper is organized as follows. In the rest part of this section, we present some preliminaries regarding encryption schemes in the cryptography field. Section~\ref{sec:pp-f} provides the system model, proposes the secure multi-party filtering protocol, and lists the problems of interests. Section~\ref{sec:converge} and Section~\ref{sec:optimal} focus on the main results including the convergence and the MMSE optimality analysis. Section~\ref{sec:numerical} provides simulations and interpretations. Section~\ref{sec:con} draws conclusions.

\emph{Notations}: 
$\mathbb{R}^n$ is the $n$-dimensional Euclidean space.
$\mathbb{R}^{n\times m}$ denotes the set of $n\times m$ real matrices. 
$\mathbb{S}_{+}^{n}$ ($\mathbb{S}_{++}^{n}$) is the set of $n\times n$ positive semi-definite (definite) matrices. When $X\in \mathbb{S}_{+}^{n}$ ($\mathbb{S}_{++}^{n}$) , we simply write $X\succeq0$ ($X\succ0$).
The identity matrix with size $n$ is represented by $I_n$. The superscript $^\top$, $\Tr\{\cdot\}$, $\rho(\cdot)$, $\left\Vert \cdot\right\Vert_{\text{F}}$, and $\left\Vert \cdot\right\Vert_{2}$ stand for the transpose, trace, spectral radius, Frobenius norm, and spectral norm of a matrix, respectively.
$\mathbb{E}[\cdot]$ denotes the mathematical expectation of a random variable. 
To save the writing space, some entries of the symmetric matrix are denoted by $*$ which can be recovered by the matrix symmetry.

\subsection{Preliminaries: Homomorphic Encryption}\label{sec:pre}
In this subsection, we introduce the \emph{homomorphic encryption}, which is the key supporting technique of this paper to protect the sensitive information of multiple parties.

The encryption mechanism is crucial to preserve the message privacy. Typically, in a secure message transmission, the encryption key is public which is usually broadcast to every party, and it is different from the decryption key which is only kept by the security module, secretly. Each party can use the public key, i.e., $f_e(\cdot)$, to encrypt a message, but only the security module with knowledge of the secret key, i.e., $f_d(\cdot)$, can decode the message:
\begin{align}
f_d\left(f_e\left(\pi_1\right)\right)=\pi_1
\end{align}
for some message $\pi_1$. It is not necessary for a conventional encryption mechanism to allow operations on encrypted messages without decryption. As a consequence, the parties' privacy will be sacrificed when they need to collaborate with other parties or utilize cloud services to process sensitive messages. To tackle such challenges, Rivest \emph{et. al}~\cite{rivest1978data} used the term ``homomorphism" for the first time to describe special encryption functions, which permit encrypted messages to be operated on without preliminary decryption. Inspired by~\cite{rivest1978data}, a burgeoning literature~\cite{rivest1978method,goldwasser2019probabilistic,elgamal1985public,benaloh1994dense,paillier1999public,gentry2009fully,smart2010fully,van2010fully,brakerski2011fully} is devoted to investigating the homomorphic scheme in the cryptography field.

The \emph{homomorphic encryption} is an encryption scheme which allows a third party (e.g., a cloud server) to apply certain operations on encrypted messages without decryption first. The homomorphic property means that given encryptions $f_e(\pi_1)$, $f_e(\pi_2)$ of messages $\pi_1$, $\pi_2$, and an operator $``\bullet"$, the result of the operation $``\bullet"$ based on encrypted messages, when decrypted, matches the result of the operation $``\circ"$ directly performed on the original plaintext, i.e.,
\begin{align}
f_d\left(f_e\left(\pi_1\right)\bullet f_e\left(\pi_2\right)\right)=\pi_1\circ \pi_2.
\end{align}
Based on the number of allowed operations on the encrypted messages, the \emph{homomorphic encryption} can be categorized into two main types: \emph{partially homomorphic encryption} (PHE) and \emph{fully homomorphic encryption} (FHE). Several widely used PHE schemes are RSA~\cite{rivest1978method}, GM~\cite{goldwasser2019probabilistic}, ElGamal~\cite{elgamal1985public}, Benaloh~\cite{benaloh1994dense}, Paillier~\cite{paillier1999public}, etc. For example, RSA is multiplicatively homomorphic, i.e., $\displaystyle f_d^{\textit{RSA}}\left(f_e^{\textit{RSA}}\left(\pi_1\right)\times f_e^{\textit{RSA}}\left(\pi_2\right)\right)=\pi_1\times\pi_2$, and Paillier is additively homomorphic, i.e., $\displaystyle f_d^{\textit{Paillier}}\left(f_e^{\textit{Paillier}}\left(\pi_1\right)\times f_e^{\textit{Paillier}}\left(\pi_2\right)\right)=\pi_1+\pi_2$ for nonnegative integers $\pi_1$ and $\pi_2$. FHE was first proposed by Gentry~\cite{gentry2009fully} which allows arbitrary operations on encrypted messages based on ideal lattices. Follow-up works~\cite{smart2010fully,van2010fully,brakerski2011fully} proposed new schemes to tackle its bottlenecks such as massive computational cost and complicated concepts.

In this paper, encryption schemes which are additively homomorphic are adopted to calculate the summation of messages from multiple parties without the leakage of parties' sensitive information. This privacy-preserving property is denoted by
\begin{align}
f_d\left(f_e\left(\pi_1\right)\oplus f_e\left(\pi_2\right)\right)=\pi_1+\pi_2,
\end{align}
where $``\oplus"$ is the corresponding operator on the ciphertext for certain additively \emph{homomorphic encryption}. Throughout this paper, we neglect the slight performance degradation induced by the quantization required before the encryption. Wa aim to get some inspiring insights theoretically.

\section{Secure Multi-Party State Estimation}\label{sec:pp-f}
\begin{figure}[t]
	\centering
	\includegraphics[width=0.49\textwidth]{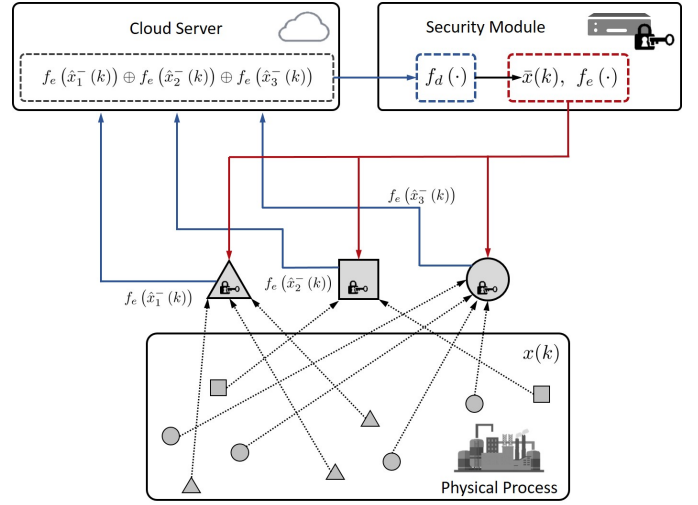}
	\caption{Diagram of secure multi-party state estimation where $N=3$. Three parties, denoted by a large triangle, square, and circle, respectively, collect measurements of the physical process via their own sensors, which are denoted by small triangles, squares, and circles, respectively.}
	\label{fig:diagram}
\end{figure}

Consider a discrete-time linear time-invariant (LTI) process in~\eqref{eqn:dynamic},
where $x\left(k\right)\in\mathbb{R}^{n}$ is state of the physical process and $w\left(k\right)\in\mathbb{R}^{n}$ is i.i.d. zero-mean Gaussian process noise with covariance $Q\succ 0$. There are $N$ parties deploying their own sensor networks to monitor the process. Each party $i$, $i\in\mathscr{N}\triangleq \{1,2,\dots,N\}$, has its own observation equation~\eqref{eqn:observation_equation}, where $C_i\in \mathbb{R}^{m_i\times n}$ is the observation matrix of party $i$, for which the concrete form is decided by the configuration of the sensors that belong to party $i$, $y_i(k)\in\mathbb{R}^{m_i}$ is the associated measurements, and $v_i(k)\in\mathbb{R}^{m_i}$ denotes i.i.d. zero-mean Gaussian measurement noise with covariance $R_i\succ 0$, which is uncorrelated with $w(k)$ and $v_j(k)$ if $j\neq i$.
The initial state $x(0)$ is zero-mean Gaussian with covariance $\Pi_0\succeq 0$, and is uncorrelated with $w(k)$ and $v_i(k)$ for all $k\geq 0$ and $i\in\mathscr{N}$.

\subsection{Secure Dynamic State Estimation Paradigm}\label{sec:filtering}
\begin{algorithm}[t]  
	\caption{Secure Multi-Party Filtering Protocol} 
	\label{alg:secure} 
	\begin{algorithmic}[1] 
		\State \textbf{Public input}: $N$;
		\State \textbf{Private input}: $y_i(k)$, $C_i$, and $K_i$ for each party $i$;
		\State \textbf{Initialization}: $\hat{x}_i(0)$ for all $i\in\mathscr{N}$;
		\State Security module: broadcast $f_e(\cdot)$ to all the $N$ parties;
		\For{$k=1,2,\dots$}
		\For{each party $i\in\mathscr{N}$} in parallel
		\State Update local state estimate $\hat{x}_i^{-}(k)$:
		\begin{align}
		\hat{x}_i^{-}(k)=A \hat{x}_i(k-1)+K_i\left(y_i(k)-C_i A\hat{x}_i(k-1)\right);\label{eqn:local}
		\end{align}
		\State Encrypt $\hat{x}_i^{-}(k)$ by $f_e(\cdot)$;
		\State Upload $f_e(\hat{x}_i^{-}(k))$ to cloud server;
		\EndFor
		\State Cloud server: perform $``\oplus"$ on received messages;
		\State Cloud server: transmit $\displaystyle f_e\left(\hat{x}_1^{-}\left(k\right)\right)\oplus\cdots\oplus f_e\left(\hat{x}_N^{-}\left(k\right)\right)$ to security module;
		\State Security module: decrypt the received message by $f_d(\cdot)$ and divide the decrypted result by $N$;
		\State Security module: broadcast $\bar{x}(k)$ to all the $N$ parties where
		\begin{align}
		\bar{x}(k)=\frac{1}{N}\sum_{i=1}^N \hat{x}_i^{-}(k);\label{eqn:average}
		\end{align}
			\For{each party $i\in\mathscr{N}$} in parallel
		\State Synchronize state estimate $\hat{x}_i(k)$ with $\bar{x}(k)$:
		\begin{align}
		\hat{x}_i(k)=\bar{x}(k);\label{eqn:fusion}
		\end{align}
		\EndFor
		\EndFor
	\end{algorithmic}  
\end{algorithm}  

We propose a secure multi-party state estimation paradigm. As depicted in Fig.~\ref{fig:diagram}, the whole filtering procedure involves a physical process, multiple parties together with their own sensors, a cloud server, and a security module. At the beginning, the public key $f_e(\cdot)$ is broadcast from the security module to all the parties. First, at every time epoch $k$, after updating the local state estimate $\hat{x}_i^{-}(k)$ based on the measurement $y_i(k)$ collected from party $i$'s own sensor network as Eq.~\eqref{eqn:local} shows, party $i$ encrypts its local state estimate $\hat{x}_i^{-}(k)$ by $f_e(\cdot)$ and uploads the encrypted message $f_e\left(\hat{x}_i^{-}\left(k\right)\right)$ to the cloud server. Second, according to the chosen encryption scheme, the cloud server performs the corresponding operator $``\oplus"$ on all the encrypted messages received and transmits the generated result $\displaystyle f_e\left(\hat{x}_1^{-}\left(k\right)\right)\oplus\cdots\oplus f_e\left(\hat{x}_N^{-}\left(k\right)\right)$ to the security module. Since the cloud server does not hold the secret key, no more information about the parties' original messages can be inferred from the encrypted messages, which demonstrates the privacy-preserving property of the \emph{homomorphic encryption}. Then the security module decrypts the received message using the secret key $f_d(\cdot)$, divides the decrypted result by the parties' total number $N$, and broadcasts it to all the parties, which is exactly the average of all the local state estimates, i.e., $\bar{x}(k)$, as shown by Eq.~\eqref{eqn:average}. Finally, each party $i$ synchronizes its state estimate $\hat{x}_i(k)$ with the average state estimate $\bar{x}(k)$ as Eq.~\eqref{eqn:fusion} shows. 

The secure multi-party state estimation paradigm is summarized in Algorithm~\ref{alg:secure}. The matrix $K_i\in\mathbb{R}^{n\times m_i}$ is the linear estimator gain of party $i$ to be decided. The initial state estimate $ \hat{x}_i(0)$ is i.i.d. zero-mean Gaussian with covariance $\widehat{\Pi}_0\succeq 0$.
During the whole filtering procedure, only the average state estimate is accessible to each party $i$, and in general, no more knowledge about the other parties' sensitive information including their observation parameters (i.e., $C_j$ and $R_j$) and their local state estimates $\hat{x}_j^{-}(k)$ where $j\in\mathscr{N}$ and $j\neq i$ can be learned, which is one of the main differences from the existing literature on the state estimation with multiple parties~\cite{shi2014event,song2015multi,yang2017multi,ribeiro2006soi,olfati2007distributed,carli2008distributed,cattivelli2010diffusion,battistelli2014consensus}. For example, when the information filter is adopted for multi-party state estimation, the unexpected disclosure of encrypted messages may reveal the estimation quality of the party, i.e., a larger $C_i^\top R_i^{-1} C_i$ implies a better estimation performance.
As a contrast, for our proposed protocol, even when the cloud server and security module conspire to get access to the data of some parties, the model privacy is still preserved.

\subsection{Problems of Interests}
In the multi-party state estimation paradigm, the local state estimates from all the parties need to be averaged and the additively \emph{homomorphic encryption} is introduced to preserve each party's data privacy. To run the filtering protocol, each party's local estimator gain $K_i$ needs to be carefully designed.

Based on standard results in control theory, it is necessary to have the knowledge of $C_i$'s, and $R_i$'s in some particular settings, to design $K_i$'s. If we design $K_i$'s in a centralized manner, each party should share its own $C_i$ (and $R_i$) and inevitably, it leads to model privacy loss. Therefore, the first and most crucial problem that we need to address is how to design $K_i$'s to stabilize the linear filtering system without disclosing each party's model privacy. Moreover, along this line of research, we are interested in finding a design method to achieve the MMSE multi-party dynamic state estimation. We hope that our findings on performance analysis can serve as motivations for multiple parties to work together to bring the proposed paradigm into life.

The performance of the filtering protocol depends on the state estimation error and its corresponding error covariance. The state estimation error covariance for party $i$ is defined as
\begin{align}
P^i(k)\triangleq \mathbb{E}\left[\left(x\left(k\right)-\hat{x}_i\left(k\right)\right)\left(x\left(k\right)-\hat{x}_i\left(k\right)\right)^\top\right].
\end{align}
Intuitively, a smaller error covariance implies a higher state estimation quality on average. After Eq.~\eqref{eqn:fusion}, all of the parties' state estimates are synchronized with $\bar{x}(k)$ and become the same, and the average state estimate $\bar{x}(k)$ is updated as:
\begin{align}
\bar{x}(k)&=\frac{1}{N}\sum_{i=1}^N\bigg{[}\left(A-K_iC_iA\right)\bar{x}(k-1)+K_i C_i A x(k-1)\nonumber\\
&~~~~~~~~~~~~~+K_iC_iw(k-1)+K_i v_i(k)\bigg{]}.\label{eqn:barx}
\end{align}
The state estimation error $\displaystyle \bar{e}(k)\triangleq x(k)-\bar{x}(k)$ is denoted by
\begin{align}
\bar{e}(k)&=\frac{1}{N}\sum_{i=1}^N \bigg{[}\left(A-K_iC_iA\right)\bar{e}\left(k-1\right)\nonumber\\
&~~~~~~~~~~~~+\left(I_n-K_iC_i\right)w\left(k-1\right)-K_iv_i\left(k\right)\bigg{]},\label{eqn:bare}
\end{align}
where $\displaystyle \mathbb{E}\left[\bar{e}\left(k\right)\right]=0$.
Its corresponding error covariance $\overline{P}(k)$, which is defined as $\displaystyle \overline{P}(k)\triangleq \mathbb{E}\left[\bar{e}\left(k\right)\bar{e}\left(k\right)^\top\right]$ and also equals to $P^i(k)$ for all $i\in\mathscr{N}$, has the following dynamic:
\begin{align}
\overline{P}(k)&=\frac{1}{N^2} \left[\sum_{i=1}^N\left(I_n-K_iC_i\right)\right]\left(A\overline{P}(k-1)A^\top+Q\right)\nonumber\\
&~~~\times \left[\sum_{i=1}^N\left(I_n-K_iC_i\right)\right]^\top+\frac{1}{N^2}\left(\sum_{i=1}^N K_i R_i K_i^\top\right)\nonumber\\
&=\frac{1}{N^2}\left(\sum_{i=1}^N M_i\right)\left(A\overline{P}(k-1)A^\top+Q\right)\left(\sum_{i=1}^N M_i^\top\right)\nonumber\\
&~~~+\frac{1}{N^2}\left(\sum_{i=1}^N S_i\right).\label{eqn:overlineP}
\end{align}
For notational simplicity, we denote $\displaystyle M_i\triangleq I_n-K_iC_i$ and $\displaystyle S_i\triangleq K_i R_i K_i^\top\succeq 0$ for all $i\in\mathscr{N}$. 

In the following two sections, the error covariance $\overline{P}(k)$ is adopted as a metric to evaluate the stability and optimality of our proposed secure multi-party state estimator.

\section{Stabilization of Multi-Party Dynamic State Estimator with Privacy Awareness}\label{sec:converge}
In this section, we analyze the stability of the error covariance $\overline{P}(k)$. First, an assumption is provided to guarantee the existence of parties' estimator gains which stabilize the estimator. Second, two methods are provided on how to choose these linear estimator gains collaboratively without leakage of model privacy.

According to Eq.~\eqref{eqn:overlineP}, the convergence of the state estimation error covariance $\overline{P}(k)$ is determined by the spectral radius of $\frac{1}{N}\sum_{i=1}^N M_iA$. Based on standard results on the discrete Lyapunov equation, it can be concluded that if and only if
\begin{align}
\rho\left(\frac{1}{N}\sum_{i=1}^N M_i A\right)<1, \label{eqn:rho_ine}
\end{align}
$\overline{P}(k)$ converges to a unique positive semi-definite value. To ensure the convergence of $\overline{P}(k)$, each party should carefully design its local estimator's linear gain $K_i$ such that the matrix $\frac{1}{N}\sum_{i=1}^N M_iA$ is stable. The following assumption provides a sufficient and necessary condition for the existence of such gain $K_i$'s. For notational convenience, we define
\begin{align}\label{eqn:C}
C\triangleq\begin{bmatrix}
C_1^\top&C_2^\top&\cdots&C_N^\top
\end{bmatrix}^\top.
\end{align}

\begin{algorithm}[t] 
	\caption{Stabilization Design Method I} 
	\label{alg:gain} 
	\begin{algorithmic}[1] 
		\State \textbf{Public input}: $N$;
		\State \textbf{Private input}: $C_i$ for each party $i$;
		\State Security module: broadcast $f_e(\cdot)$ to all the $N$ parties;
		\For{each party $i\in\mathscr{N}$} in parallel
		\State Calculate estimator gain $K_i$:
		\begin{align}
		K_i=\argmin_{X\in\mathbb{R}^{n\times m_i}} \left\Vert \left(I_n-X C_i\right)A\right\Vert;
		\end{align}
		\State Encrypt $\left\Vert \left(I_n-K_i C_i\right)A\right\Vert$ by $f_e(\cdot)$;
		\State Upload $f_e\left(\left\Vert \left(I_n-K_i C_i\right)A\right\Vert\right)$ to cloud server;
		\EndFor
		\State Cloud server: perform $``\oplus"$ on received messages;
		\State Cloud server: transmit $\displaystyle f_e\left(\left\Vert \left(I_n-K_1 C_1\right)A\right\Vert\right)\oplus\cdots\oplus f_e\left(\left\Vert \left(I_n-K_N C_N\right)A\right\Vert\right)$ to security module;
		\State Security module: decrypt the received message by $f_d(\cdot)$ and divide the decrypted result by $N$;
		\State Security module: tell whether 
		\begin{align}\label{eqn:argmin}
		\frac{1}{N} \sum_{i=1}^N \left\Vert\left(I_n-K_iC_i\right)A\right\Vert<1
		\end{align}
		and broadcast the result to all the $N$ parties.
	\end{algorithmic}  
\end{algorithm} 

\begin{assumption}\label{asmp:detec}
	The pair $(A, C)$ is detectable.
\end{assumption}

\begin{remark}\label{lem:equivalent}
The pair $(A, C)$ being detectable is equivalent to $(A, CA)$ being detectable. To see this, first, since $(A, CA)$ is detectable, one can always find a suitable $K$ such that $A-KCA$ is stable. Note that $(I_n-KC)A$ and $A(I_n-KC)$ have the same eigenvalues, and therefore $A-AKC$ is also stable. Obviously, the pair $(A, C)$ is detectable. Conversely, from~\cite{kailath2000linear}, if $(A, C)$ is detectable, then for any $\widehat{R}\succ 0$ and $\widehat{Q}\succeq 0$ satisfying that $(A, \sqrt{\widehat{Q}})$ is stabilizable, the discrete-time algebraic Riccati equation (DARE) $X=AXA^\top+\widehat{Q}-AXC^\top\left(CXC^\top+\widehat{R}\right)^{-1}CXA^\top$ has a unique solution $\widehat{X}\succeq 0$. Let $K\triangleq \widehat{X}C^\top\left(C\widehat{X}C^\top+\widehat{R}\right)^{-1}$, and one can conclude that $A-KCA$ is stable according to the standard Kalman filter results. By this construction method, we find a suitable $K$ to make $A-KCA$ stable and therefore, $(A, CA)$ is detectable.
\end{remark}

Assumption~\ref{asmp:detec} is arguably the weakest assumption in the context of dynamic state estimation. Otherwise, all the parties cannot even achieve a stable estimator under a centralized setup. Remark~\ref{lem:equivalent} suggests that Assumption~\ref{asmp:detec} is equivalent to the detectability of $(A, CA)$. When $(A,C)$ is detectable, we can always find a matrix $K$, where $K=\frac{1}{N}\begin{bmatrix}
K_1&K_2&\cdots&K_N
\end{bmatrix}$ with each $K_i$ being of proper dimension in accordance with $C_i$, that makes $A-KCA$ stable.
These $K_i$'s guarantee the feasibility of~\eqref{eqn:rho_ine}. 

In the following, we will develop two collaboratively design methods to stabilize the multi-party state estimator without disclosing sensitive model information (i.e., the observation matrix $C_i$) for the sake of protecting model privacy. 

\subsection{Stabilization Design Method I: Matrix Norm Relaxation}\label{sec:matrix_norm} 
The first design method (Algorithm~\ref{alg:gain}) utilizes the matrix norm relaxation and additively \emph{homomorphic encryption}. For any induced matrix norm $\Vert \cdot \Vert$, there holds:
\begin{align}
\rho\left(\frac{1}{N}\sum_{i=1}^N M_i A\right)\leq \left\Vert \frac{1}{N}\sum_{i=1}^N M_i A\right\Vert\leq \frac{1}{N}\sum_{i=1}^N \left\Vert M_i A\right\Vert.
\end{align}
Note that the local estimator gain $K_i$ could be determined such that $\displaystyle \left\Vert M_i A\right\Vert<1$ for each party $i$, then~\eqref{eqn:rho_ine} holds trivially and  $\overline{P}(k)$ converges. However, the condition $\displaystyle \left\Vert M_i A\right\Vert<1,~\forall~i\in\mathscr{N}$ restricts the choice of the linear gain $K_i$, and sometimes such a $K_i$ may even not exist. We propose a simple design method in Algorithm~\ref{alg:gain}. This algorithm is primitive and optional. Similarly to Algorithm~\ref{alg:secure} in Section~\ref{sec:filtering}, after each party decides its estimator gain $K_i$, the average $\frac{1}{N}\sum_{i=1}^N \left\Vert M_i A\right\Vert$ can be calculated with the help of the cloud server and the security module, which does not leak any private $\displaystyle \left\Vert M_i A\right\Vert$. If the average is less than $1$, then the candidate estimator gains are acceptable and a bounded estimation error covariance is guaranteed. 
This method is applicable when $\displaystyle \min_{\{K_i\}}~\frac{1}{N} \sum_{i=1}^N \left\Vert\left(I_n-K_iC_i\right)A\right\Vert<1$.

\subsection{Stabilization Design Method II: ADMM}\label{sec:ADMM}
\begin{algorithm*}[t]
	\caption{Stabilization Design Method II}
	\label{alg:ADMM}
	\begin{algorithmic}[1]
		\State \textbf{Public input}: $N,~\gamma$;
		\State \textbf{Private input}: $C_i$ for each party $i$;
		\State \textbf{Initialization}: $Z_i^0$ and $\Lambda_i^0$ for all $i\in\mathscr{N}$;
		\For{iteration $t=1,2,\dots$}
		\State Cloud server: update $\big{(}\left\{U_i\right\},~\bm{U_0},~H\big{)}$ by solving the convex subproblem:
		\begin{align*}
		\left(\left\{U_i^{t}\right\}, \bm{U_0}^{t}, H^{t}\right)=&\argmin_{\left(\left\{U_i\right\}, \bm{U_0}, H\right)}~~-\sum_{i=1}^N \Tr\left\{\left(\Lambda_i^{t-1}\right)^\top\left(Z_i^{t-1} C_i A-U_i\right) \right\}+\frac{\gamma}{2}\sum_{i=1}^N \left\Vert Z_i^{t-1} C_iA-U_i \right\Vert_{\text{F}}^2,\\ &~~~~~\text{s.t.}~~~~~~~~~\big{(}\left\{U_i\right\}, \bm{U_0}, H\big{)}\in\Omega;
		\end{align*}
		\State Cloud server: feedback $\left(U_i^{t},~\Lambda_i^{t-1}\right)$ to each party $i$;
		\For{each party $i\in\mathscr{N}$} in parallel
		\State Update $Z_i$ by solving the convex subproblem:
		\begin{align*}
		Z_i^{t}=\argmin_{Z_i}~~-\Tr\left\{\left(\Lambda_i^{t-1}\right)^\top\left(Z_i C_i A-U_i^{t}\right) \right\}+\frac{\gamma}{2} \left\Vert Z_i C_iA-U_i^{t} \right\Vert_{\text{F}}^2;
		\end{align*}
		\State Upload $Z_i^{t} C_i A$ to cloud server;
		\EndFor		
		\State Cloud server: update $\displaystyle \Lambda_i^{t}=\Lambda_i^{t-1}-\gamma \left(Z_i^{t} C_i A-U_i^{t}\right)$ for all $i\in\mathscr{N}$;
		\EndFor
		\State Cloud server: broadcast $H^t$ to all the $N$ parties;
		\State \textbf{Output}: Linear estimator gain $K_i^t=\left(H^t\right)^{-1} Z_i^t$ for all $i\in\mathscr{N}$.		
	\end{algorithmic}
\end{algorithm*}

The second design method (Algorithm~\ref{alg:ADMM}), which solves a formulated convex semi-definite programming problem by ADMM, is applicable to general cases under Assumption~\ref{asmp:detec}. The design purpose is to find the estimator gain $K_i$'s such that $A-KCA$ is stable, where
\begin{align}\label{eqn:K}
K\triangleq\frac{1}{N}\begin{bmatrix}
K_1&K_2&\cdots&K_N
\end{bmatrix}
\end{align}
and $C$ is as defined in Eq.~\eqref{eqn:C}. For notational convenience, $m\triangleq \sum_{i=1}^N m_i$. Basically, how to find a $K$ to stabilize $A-KCA$ has already been well investigated in the control field, for example, the pole placement problem~\cite{chen1998linear}. However, parties' concerns about their sensitive information, e.g., $C_i$, prohibit a centralized design. To disclose as little sensitive information as possible, we formulate an equivalent convex problem and solve it by ADMM.

First, we transform the collaborative gain design problem into an optimization problem.
The stabilization of $A-KCA$ is equivalent to the following condition according to the standard results on the Lyapunov equation
\begin{align}
&\exists~\Delta\in \mathbb{S}_{++}^{n}~\text{and}~K\in\mathbb{R}^{n\times m},\nonumber\\ &\text{s.t.}~\Delta-\left(A-KCA\right)\Delta\left(A-KCA\right)^\top\succ 0,\label{eqn:lya}
\end{align}
which is further equivalent to
\begin{align}
&\exists~\Delta\in \mathbb{S}_{++}^{n}~\text{and}~K\in\mathbb{R}^{n\times m},\nonumber\\
&\text{s.t.}~\begin{bmatrix}
\Delta&A-KCA\\
\left(A-KCA\right)^\top&\Delta^{-1}
\end{bmatrix}\succ 0 \label{eqn:bothsides}
\end{align}
due to the Schur complement condition for positive definiteness. By multiplying $\displaystyle \begin{bmatrix}
\Delta^{-1}&\\
&I_n
\end{bmatrix}$ on both sides of~\eqref{eqn:bothsides}, we have the following equivalent condition:
\begin{align*}
&\exists~\Delta\in \mathbb{S}_{++}^{n}~\text{and}~K\in\mathbb{R}^{n\times m},\nonumber\\
&\text{s.t.}~\begin{bmatrix}
\Delta^{-1}&\Delta^{-1}A-\Delta^{-1}KCA\\
\left(\Delta^{-1}A-\Delta^{-1}KCA\right)^\top&\Delta^{-1}
\end{bmatrix}\succ 0.
\end{align*}
For notational brevity, we denote $H\triangleq \Delta^{-1}$ and $Z_i\triangleq \Delta^{-1}K_i$. By algebraic manipulation and change of variables, the problem to find suitable $H$ and $Z_i$'s is equivalent to an optimization problem as follows.

\begin{problem}\label{prob:ADMM}
	\begin{align*}
	\min_{\{Z_i\},H}~~~~~~&\text{constant},\\
	\text{s.t.}~~~~~~~&\sum_{i=1}^N \begin{bmatrix}
	H&HA-Z_iC_iA\\
	(HA-Z_iC_iA)^\top&H
	\end{bmatrix}\succ 0,\\
	&H\succ 0.
	\end{align*}
\end{problem}

The ADMM algorithm is remarkably successful in solving convex programs which minimize a sum of $N$ convex objectives whose variables are linked by some constraints. Although the convergence of ADMM for $N=2$ is a standard result, the result for scenario where $N\geq 3$ is not that obvious. In particular, it was shown by Chen \emph{et al.}~\cite{chen2016direct} that ADMM for $N\geq 3$ fails to converge in general. To convert the multi-block problem into an equivalent two-block problem, we introduce an indicator function
\begin{align}
\mathcal{I}_\Omega\left(\left\{U_i\right\}, \bm{U_0}, H\right)=\begin{cases}
~~0,&\text{if}~\big{(}\left\{U_i\right\}, \bm{U_0}, H\big{)}\in\Omega,\\
+\infty,&\text{otherwise},
\end{cases}
\end{align}
where the convex set $\Omega$ is given by
\begin{align}
\Omega\triangleq \bigg{\{} &\big{(}\left\{U_i\right\}, \bm{U_0}, H\big{)}: \bm{U_0}\succ 0,~~H\succ 0,\nonumber \\
&N\begin{bmatrix}
H&HA\\
(HA)^\top&H
\end{bmatrix}-\sum_{i=1}^N \begin{bmatrix}
0&U_i\\
U_i^\top&0
\end{bmatrix}=\bm{U_0} \bigg{\}}
\end{align}
for $U_i\in\mathbb{R}^{n\times n}$, $i\in\mathscr{N}$. By variable splitting, an equivalent two-block problem is formulated as follows.

\begin{problem}\label{prob:ADMM_variant}
	\begin{align*}
	\min_{\left\{Z_i\right\},\left\{U_i\right\}, \bm{U_0}, H}~~~~~~&\mathcal{I}_\Omega\left(\left\{U_i\right\}, \bm{U_0}, H\right),\\
	\text{s.t.}~~~~~~~~~~~~~&Z_i C_iA=U_i,~\forall~i\in\mathscr{N}.
	\end{align*}
\end{problem}
The variables in Problem~\ref{prob:ADMM_variant} can be grouped into two blocks: $\big{(}\left\{U_i\right\},~\bm{U_0},~H\big{)}$ and $\{Z_i\}$, so that ADMM can directly apply. The augmented Lagrangian where $\Lambda_i\in\mathbb{R}^{n\times n}$ is the corresponding Lagrange multiplier for Problem~\ref{prob:ADMM_variant} is given by:
\begin{align*}
&\mathcal{L}_\gamma\left(\left\{Z_i\right\}, \left\{U_i\right\}, \bm{U_0}, H, \left\{\Lambda_i\right\}\right)=\mathcal{I}_\Omega\left(\left\{U_i\right\}, \bm{U_0}, H\right)\\
&~~-\sum_{i=1}^N \Tr\left\{\Lambda_i^\top\left(Z_iC_i A-U_i\right) \right\}+\frac{\gamma}{2}\sum_{i=1}^N \left\Vert Z_iC_iA-U_i \right\Vert_{\text{F}}^2,
\end{align*}
which incorporates a quadratic penalty of the constraint scaled by a parameter $\gamma>0$ into the Lagrangian. Since $Z_i$'s are fully decoupled, the resulting subproblem is decomposed into $N$ separated subproblems, which can be solved in parallel. The details are summarized in Algorithm~\ref{alg:ADMM}.

Recall that we aim to collaboratively design the linear estimator gains without exposing sensitive model information such that the state estimation error covariance $\overline{P}(k)$ converges. By adopting Algorithm~\ref{alg:ADMM}, all the parties compute their gains together with the cloud server. First, the cloud server updates the variables  $\big{(}\left\{U_i^t\right\},~\bm{U_0}^t,~H^t\big{)}$ and feedbacks $\left(U_i^t,~\Lambda_i^{t-1}\right)$ to each party $i$, respectively. And then each party updates its own $Z_i^t$ in parallel. Note that at each iteration $t$, party $i$ can upload the value of $Z_i^t C_i A$ to the cloud server. There is no need for party $i$ to expose its own observation matrix $C_i$ to others. In other words, the sensitive information of all parties is protected by the designed method. After the convergence of Algorithm~\ref{alg:ADMM}, each party obtains its estimator gain by $K_i=H^{-1} Z_i$.

Whenever there is a new party, termed as the $(N+1)$-th party, to join the protocol, it can simply set $\displaystyle K_{N+1}=\argmin_{X\in\mathbb{R}^{n\times m_{N+1}}} \left\Vert \left(I_n-X C_{N+1}\right)A\right\Vert$. If the existing $N$ parties have decided gains by Algorithm~\ref{alg:ADMM}, now the $N+1$ parties need to check whether $\frac{1}{N+1}\left(\left\Vert\sum_{i=1}^{N} M_{i}A \right\Vert+\left\Vert M_{N+1}A \right\Vert\right)<1$. If the existing $N$ parties have decided gains by Algorithm~\ref{alg:gain}, now the $N+1$ parties need to check whether the resulting new average $\frac{1}{N+1}\sum_{i=1}^{N+1} \left\Vert M_i A\right\Vert<1$. If the summation is less than $1$, then the new party is allowed to join based on the results derived in Section~\ref{sec:matrix_norm}. Otherwise, all the $N+1$ parties need to collaboratively design the new stable estimator gains by performing Algorithm~\ref{alg:ADMM}.

\begin{remark}
	Both the above design methods guarantee the stability of the proposed filtering protocol. By the two design methods, it is also possible to assign the convergence rate of the overall state estimation to an arbitrary level. If we prescribe the convergence rate $\rho\left(\frac{1}{N}\sum_{i=1}^N M_i A \right)$ to be less than $\epsilon$, where $\epsilon\in (0,1]$, we could use $\frac{1}{\epsilon}A$ to replace $A$ in the first and second stabilization design methods. The rest of Algorithm~\ref{alg:gain} and~\ref{alg:ADMM} remains unchanged.
\end{remark}

\section{Asymptotic MMSE Secure Multi-Party State Estimation}\label{sec:optimal}
\begin{algorithm*}[t]
	\caption{Asymptotic MMSE Gain Design Method with Privacy Guarantees}
	\label{alg:ADMM_2}
	\begin{algorithmic}[1]
		\State \textbf{Public input}: $N,~\gamma$, $m_i$ for all $i\in\mathscr{N}$;
		\State \textbf{Private input}: $C_i$ and $R_i$ for each party $i$;
		\State \textbf{Initialization}: $\overline{Z}_i^0$ and $\overline{\Lambda}_i^0$ for all $i\in\mathscr{N}$;
		\For{iteration $t=1,2,\dots$}
		\State Cloud server: update $\left(\left\{\overline{U}_i\right\},\left\{\overline{V}_i\right\}, \bm{\overline{W}}, \overline{H} \right)$ by solving the convex subproblem:
		\begin{align*}
		\left(\left\{\overline{U}_i^t\right\},\left\{\overline{V}_i^t\right\}, \bm{\overline{W}}^t, \overline{H}^t \right)=\argmin_{\left(\left\{\overline{U}_i\right\},\left\{\overline{V}_i\right\}, \bm{\overline{W}}, \overline{H} \right)}~~&-\Tr\{\overline{H}\}-\sum_{i=1}^N \Tr\left\{\left(\overline{\Lambda}_i^{t-1}\right)^\top\left(\overline{Z}_i^{t-1}\begin{bmatrix}
		C_i&\sqrt{R_i}B_i
		\end{bmatrix}-\begin{bmatrix}
		\overline{U}_i&\overline{V}_i
		\end{bmatrix}\right) \right\}\\
		&+\frac{\gamma}{2}\sum_{i=1}^N \left\Vert \overline{Z}_i^{t-1}\begin{bmatrix}
		C_i&\sqrt{R_i}B_i
		\end{bmatrix}-\begin{bmatrix}
		\overline{U}_i&\overline{V}_i
		\end{bmatrix} \right\Vert_{\text{F}}^2,\\
		\text{s.t.}~~~~~~~~~~&\left(\left\{\overline{U}_i\right\},\left\{\overline{V}_i\right\}, \bm{\overline{W}}, \overline{H} \right)\in\overline{\Omega};
		\end{align*}
		\State Cloud server: feedback $\left(\overline{U}_i^t,\overline{V}_i^t, \overline{\Lambda}_i^{t-1}\right)$ to each party $i$;
		\For{each party $i\in\mathscr{N}$} in parallel
		\State Update $\overline{Z}_i$ by solving the convex subproblem:
		\begin{align*}
		\overline{Z}_i^{t}=\argmin_{\overline{Z}_i}~-\Tr\left\{\left(\overline{\Lambda}_i^{t-1}\right)^\top\left(\overline{Z}_i\begin{bmatrix}
		C_i&\sqrt{R_i}B_i
		\end{bmatrix}-\begin{bmatrix}
		\overline{U}_i^t&\overline{V}_i^t
		\end{bmatrix}\right) \right\}+\frac{\gamma}{2} \left\Vert \overline{Z}_i\begin{bmatrix}
		C_i&\sqrt{R_i}B_i
		\end{bmatrix}-\begin{bmatrix}
		\overline{U}_i^t&\overline{V}_i^t
		\end{bmatrix} \right\Vert_{\text{F}}^2;
		\end{align*}
		\State Upload $\overline{Z}_i^{t} C_i$ and $\overline{Z}_i^t \sqrt{R_i}B_i$ to cloud server;
		\EndFor		
		\State Cloud server: update $\displaystyle \overline{\Lambda}_i^{t}=\overline{\Lambda}_i^{t-1}-\gamma \left(\overline{Z}_i^{t}\begin{bmatrix}
		C_i&\sqrt{R_i}B_i
		\end{bmatrix}-\begin{bmatrix}
		\overline{U}_i^t&\overline{V}_i^t
		\end{bmatrix}\right)$ for all $i\in\mathscr{N}$;
		\EndFor
		\State Cloud server: broadcast $\overline{H}^t$ to all the $N$ parties;
		\State \textbf{Output}: Linear estimator gain $K_i^t=A^{-1}\left(\overline{H}^t\right)^{-1} \overline{Z}_i^t$ for all $i\in\mathscr{N}$.		
	\end{algorithmic}
\end{algorithm*}

When statistics of the noise process are known \emph{a priori}, the MMSE estimator, which minimizes a quadratic error cost function in a Bayesian setting, is recognized as a popular choice. In this section, we will discuss MMSE secure multi-party dynamic state estimation design that preserves model privacy. 
Recall that by adopting the proposed secure multi-party dynamic state estimation paradigm, the average state estimate $\bar{x}(k)$ (see Eq.~\eqref{eqn:barx}) evolves as: 
\begin{align}
\bar{x}(k)=A\bar{x}(k-1)+K\left(y\left(k\right)-CA\bar{x}\left(k-1\right)\right),
\end{align}
where $y(k)\triangleq\begin{bmatrix}
y_1^\top(k)&y_2^\top(k)&\cdots&y_N^\top(k)
\end{bmatrix}^\top$, $C$ and $K$ are as defined in Eq.~\eqref{eqn:C} and~\eqref{eqn:K}, respectively. 
If the pair $(A, C)$ is detectable and $Q\succ 0$, then the optimal steady-state Kalman gain~\cite{kailath2000linear} asymptotically achieves MMSE estimation and is given by $K^\star=P_{\text{pri}}C^\top\left(CP_{\text{pri}}C^\top+R\right)^{-1}$, where $P_{\text{pri}}\succ 0$ is the solution to the DARE $\tilde{g}(X)=X$ for $\tilde{g}(X)\triangleq AXA^\top +Q-AXC^\top\left(CXC^\top+R \right)^{-1}CXA^\top$. Here, $\displaystyle R\triangleq \text{diag}\left\{R_1,R_2,\dots,R_N\right\}$. In this section, we target at obtaining the optimal steady-state Kalman gain $K^\star$ while protecting the observation parameters $C_i$ and $R_i$ of party $i$. To establish this, we need the following Lemma~\ref{lem:bruno} as preliminaries. 
The remaining results in this section are developed under Assumption~\ref{asmp:nonsingular}.
\begin{assumption}\label{asmp:nonsingular}
	The pair $(A,C)$ is detectable and $A$ is nonsingular.
\end{assumption}

\begin{remark}
	Note that the process dynamic matrix $A$ is always nonsingular if the LTI process~\eqref{eqn:dynamic} is a discretization from a continuous-time state space model, i.e., $\displaystyle \dot{x}_c(t)=A_c x_c(t)+w_c(t)$. One can check that the dynamic matrix of the corresponding discretized model is $A=e^{A_c T_s}$ where $T_s$ is the sampling time~\cite{chen1998linear}, and $A$ must be nonsingular.
\end{remark}

For notational convenience, the operator $\phi(\cdot,\cdot): \mathbb{R}^{n\times m}\times \mathbb{S}^n_{+}\to \mathbb{S}^n_{+}$ is defined as:
\begin{align*}
\phi(Y,X)=\left(A-AYC \right)X\left(A-AYC \right)^\top+Q+AYRY^\top A^\top.
\end{align*}

\begin{lemma}\label{lem:bruno}
	Under Assumption~\ref{asmp:nonsingular}, the following statements are equivalent.
	\begin{enumerate}
		\item $\exists~\widetilde{K}\in\mathbb{R}^{n\times m},~\widetilde{X}\in\mathbb{S}^{n}_{++}$ such that $\widetilde{X}\succeq \phi\left(\widetilde{K},\widetilde{X} \right)$;
		\item $\exists~\widetilde{\Upsilon}\in\mathbb{R}^{n\times m},~\widetilde{\Delta}\in\mathbb{S}^{n}_{++}$ such that
		\begin{align*}
		\begin{bmatrix}
		\widetilde{\Delta}&\widetilde{\Delta}A-\widetilde{\Upsilon}C&\widetilde{\Upsilon}\sqrt{R}&\widetilde{\Delta}\sqrt{Q}\\
		A^\top\widetilde{\Delta}-C^\top\widetilde{\Upsilon}^\top&\widetilde{\Delta}&0&0\\
		\sqrt{R}\widetilde{\Upsilon}^\top&0&I_{m}&0\\
		\sqrt{Q}\widetilde{\Delta}&0&0&I_n
		\end{bmatrix}\succeq 0.
		\end{align*} 
	\end{enumerate}
\end{lemma} 

\begin{IEEEproof}
Similar to the proof of Theorem 5 in~\cite{sinopoli2004kalman}, based on the Schur complement condition for positive semi-definiteness, statement $1)$ is equivalent to $\exists~\widetilde{K}\in\mathbb{R}^{n\times m},~\widetilde{X}\in\mathbb{S}^{n}_{++}$ s.t.
\begin{align}\label{eqn:tildeX}
\begin{bmatrix}
\widetilde{X}&A-A\widetilde{K}C&A\widetilde{K}\sqrt{R}&\sqrt{Q}\\
(A-A\widetilde{K}C)^\top&\widetilde{X}^{-1}&0&0\\
\sqrt{R}\widetilde{K}^\top A^\top&0&I_{m}&0\\
\sqrt{Q}&0&0&I_n
\end{bmatrix}\succeq 0.
\end{align}
By multiplying $\begin{bmatrix}
\widetilde{X}^{-1}&0\\
0&I_{2n+m}
\end{bmatrix}$ on both sides of~\eqref{eqn:tildeX} and let $\widetilde{\Delta}=\widetilde{X}^{-1}, \widetilde{\Upsilon}=\widetilde{X}^{-1}A\widetilde{K}$, statement $1)$ is equivalent to statement $2)$.	
\end{IEEEproof}

We propose the following semi-definite programming Problem~\ref{prob:optimal}, and prove that its optimal solution gives the optimal steady-state Kalman gain $K^\star$ in Theorem~\ref{thm:optimal}. For notational brevity, we denote $B_1\triangleq \begin{bmatrix}
I_{m_1}&0
\end{bmatrix}\in\mathbb{R}^{m_1\times m}$, and $B_i\triangleq [\underbrace{0~\cdots~0}_{\sum_{j=1}^{i-1} m_j}~I_{m_i}~0]\in\mathbb{R}^{m_i\times m}$ for $i\geq 2$.  

\begin{problem}\label{prob:optimal}
	\begin{align*}
	\min_{\left\{\overline{Z}_i\right\}, \overline{H}}~~~&-\Tr\{\overline{H}\},\\
	\text{s.t.}~~~~~~&\overline{H}\succ 0,\\
	&\sum_{i=1}^N\begin{bmatrix}
	\overline{H}&*&*&*\\
	A^\top\overline{H}-C_i^\top\overline{Z}_i^\top&\overline{H}&0&0\\
	B_i^\top \sqrt{R_i}\overline{Z}_i^\top&0&I_m&0\\
	\sqrt{Q}\overline{H}&0&0&I_n
	\end{bmatrix}\succeq 0.
	\end{align*}
\end{problem}

\begin{theorem}~\label{thm:optimal}
	The optimal solution to Problem~\ref{prob:optimal}, i.e., $\left\{\overline{Z}_i^\star \right\}$ and $\overline{H}^\star$, gives the optimal steady-state Kalman gain $K^\star=\frac{1}{N}A^{-1}\left(\overline{H}^\star\right)^{-1} \begin{bmatrix}
	\overline{Z}_1^\star&\overline{Z}_2^\star&\cdots&\overline{Z}_N^\star
	\end{bmatrix}$ and the fixed point of the DARE, i.e., $P_{\text{pri}}=\left(\overline{H}^\star\right)^{-1}$.
\end{theorem}

\begin{IEEEproof}
	We set the estimator gain $K_i=A^{-1}\overline{H}^{-1}\overline{Z}_i$, $K=\frac{1}{N}\begin{bmatrix}
	K_1&K_2&\cdots&K_N
	\end{bmatrix}$, and let $P=\overline{H}^{-1}$. The constraints in Problem~\ref{prob:optimal} are equivalent to $P\succ 0$ and $P\succeq \phi(K,P)$ according to Lemma~\ref{lem:bruno}. Clearly, the optimal steady-state Kalman gain $K^\star$ and the positive definite solution $P_{\text{pri}}$ to the DARE satisfy $P_{\text{pri}}=\phi(K^\star,P_{\text{pri}})$, and thus belong to the feasible set of the optimization problem. 
	
	We now prove Theorem~\ref{thm:optimal} by contradiction in two steps. First, we prove that the optimal solution reaches the equality $P=\phi(K,P)$.
	Suppose that $\widehat{K}$ and $\widehat{P}$ solve Problem~\ref{prob:optimal} but $\widehat{P}\neq \phi(\widehat{K},\widehat{P})$. Then there exists $\widehat{P}\succeq \phi(\widehat{K},\widehat{P})\succeq \tilde{g}(\widehat{P})\succ 0$ but $\Tr\{\widehat{P} \}>\Tr\{\phi(\widehat{K},\widehat{P}) \}$. The inequality $\phi(\widehat{K},\widehat{P})\succeq \tilde{g}(\widehat{P})$ is due to $\tilde{g}(X)=\min_Y~\phi(Y,X)\preceq \phi(Y,X),~\forall~Y\in\mathbb{R}^{n\times m}$ (see Lemma 1 in~\cite{sinopoli2004kalman}). We denote $\check{P}\triangleq \tilde{g}(\widehat{P})$. By setting $\check{K}=\check{P}C^\top\left(C\check{P}C^\top+R \right)^{-1}$, we have $\check{P}\succ 0$ and $\check{P}\succeq \phi(\check{K},\check{P})$ which satisfy the constraints. However, the inequalities $\widehat{P}\succeq \check{P}\succ 0$ and $\Tr\{\widehat{P} \}>\Tr\{\check{P} \}$ imply $-\Tr\left\{\widehat{P}^{-1} \right\}>-\Tr\left\{\check{P}^{-1} \right\}$, which contradicts the hypothesis of optimality of $\widehat{P}$. Therefore, the optimal solution to Problem~\ref{prob:optimal} must satisfy $\widehat{P}= \phi(\widehat{K},\widehat{P})$. 
	Second, we show that the optimal solution $K$ and $P$ must also satisfy $K=PC^\top\left(CPC^\top+R\right)^{-1}$, i.e., $P=\phi(K,P)=\tilde{g}(P)$. Suppose that $\widehat{K}$ and $\widehat{P}$ solve Problem~\ref{prob:optimal} but $\widehat{P}\neq \tilde{g}(\widehat{P})$. Then there exists $\widehat{P}=\phi(\widehat{K},\widehat{P})\succeq \tilde{g}(\widehat{P})\succ 0$ but $\Tr\{\widehat{P} \}>\Tr\{\tilde{g}(\widehat{P}) \}$. We denote $\check{P}\triangleq \tilde{g}(\widehat{P})$ and similar to the first step, $-\Tr\left\{\widehat{P}^{-1} \right\}>-\Tr\left\{\check{P}^{-1} \right\}$ can be shown, which again contradicts the hypothesis of optimality of $\widehat{P}$. Therefore, the optimal solution $K$ and $P$ to Problem~\ref{prob:optimal} must also satisfy $K=PC^\top\left(CPC^\top+R\right)^{-1}$, i.e., $P=\tilde{g}(P)$. In other words, the optimal Kalman gain $K^\star$ and the fixed point of the DARE $P_{\text{pri}}$ can be represented by the optimal solution to Problem~\ref{prob:optimal} and this concludes the theorem.   
\end{IEEEproof}

According to Theorem~\ref{thm:optimal}, solving Problem~\ref{prob:optimal} leads to the optimal steady-state Kalman gain which is of our interests. However, with privacy concerns, we do not directly solve it. Similar to the stabilization design method II in Section~\ref{sec:ADMM}, we introduce an indicator function

\begin{small}
\begin{align}
&\mathcal{I}_{\overline{\Omega}}\left(\left\{\overline{U}_i\right\},\left\{\overline{V}_i\right\}, \bm{\overline{W}}, \overline{H}\right)\nonumber\\
&~~~~~~~~~~=\begin{cases}
~~0,&\text{if}~\big{(}\left\{\overline{U}_i\right\},\left\{\overline{V}_i\right\}, \bm{\overline{W}}, \overline{H}\big{)}\in\overline{\Omega},\\
+\infty,&\text{otherwise},
\end{cases}
\end{align}
where the convex set $\overline{\Omega}$ is given by
\begin{align*}
&\overline{\Omega}\triangleq \Bigg{\{} \big{(}\left\{\overline{U}_i\right\},\left\{\overline{V}_i\right\}, \bm{\overline{W}}, \overline{H}\big{)}:~\bm{\overline{W}}\succeq 0,~~\overline{H}\succ 0,\nonumber \\
&N\begin{bmatrix}
\overline{H}&*&0&*\\
A^\top\overline{H}&\overline{H}&0&0\\
0&0&I_m&0\\
\sqrt{Q}\overline{H}&0&0&I_n
\end{bmatrix}+\sum_{i=1}^N \begin{bmatrix}
0&*&*&0\\
-\overline{U}_i^\top&0&0&0\\
\overline{V}_i^\top&0&0&0\\
0&0&0&0
\end{bmatrix}=\bm{\overline{W}} \Bigg{\}}
\end{align*}
\end{small}for $\overline{U}_i\in\mathbb{R}^{n\times n}$, $\overline{V}_i\in\mathbb{R}^{n\times m}$, $i\in\mathscr{N}$. By variable splitting, a two-block problem which is equivalent to Problem~\ref{prob:optimal} is formulated as follows.
\begin{problem}\label{prob:optimal_variant}
\begin{align*}
	\min_{\left\{\overline{Z}_i\right\},\left\{\overline{U}_i\right\},\left\{\overline{V}_i\right\}, \bm{\overline{W}}, \overline{H}}&-\Tr\{\overline{H}\}+\mathcal{I}_{\overline{\Omega}}\left(\left\{\overline{U}_i\right\},\left\{\overline{V}_i\right\}, \bm{\overline{W}}, \overline{H}\right),\\
	\text{s.t.}~~~~~~~\overline{Z}_i&\begin{bmatrix}
	C_i&\sqrt{R_i}B_i
	\end{bmatrix}=\begin{bmatrix}
	\overline{U}_i&\overline{V}_i
	\end{bmatrix},~\forall~i\in\mathscr{N}.
\end{align*}
\end{problem}
Similar to the procedure in Section~\ref{sec:ADMM}, the variables in Problem~\ref{prob:optimal_variant} can be grouped into two blocks: $\left(\left\{\overline{U}_i\right\},\left\{\overline{V}_i\right\}, \bm{\overline{W}}, \overline{H} \right)$ and $\left\{\overline{Z}_i \right\}$, so that ADMM can directly apply. The augmented Lagrangian for Problem~\ref{prob:optimal_variant} where $\overline{\Lambda}_i\in\mathbb{R}^{n\times (n+m)}$ is the corresponding Lagrange multiplier is given by:
\begin{align*}
&\mathcal{L}_\gamma\left(\left\{\overline{Z}_i\right\},\left\{\overline{U}_i\right\},\left\{\overline{V}_i\right\}, \bm{\overline{W}}, \overline{H}, \left\{\overline{\Lambda}_i\right\}\right)\\
=&-\Tr\{\overline{H}\}+\mathcal{I}_{\overline{\Omega}}\left(\left\{\overline{U}_i\right\},\left\{\overline{V}_i\right\}, \bm{\overline{W}}, \overline{H}\right)\\
&-\sum_{i=1}^N \Tr\left\{\overline{\Lambda}_i^\top\left(\overline{Z}_i\begin{bmatrix}
C_i&\sqrt{R_i}B_i
\end{bmatrix}-\begin{bmatrix}
\overline{U}_i&\overline{V}_i
\end{bmatrix}\right) \right\}\\
&+\frac{\gamma}{2}\sum_{i=1}^N \left\Vert \overline{Z}_i\begin{bmatrix}
C_i&\sqrt{R_i}B_i
\end{bmatrix}-\begin{bmatrix}
\overline{U}_i&\overline{V}_i
\end{bmatrix} \right\Vert_{\text{F}}^2.
\end{align*}
Since $\overline{Z}_i$'s are fully decoupled, the resulting subproblems can be solved in parallel. Algorithm~\ref{alg:ADMM_2} is proposed to solve Problem~\ref{prob:optimal_variant}. After the convergence of the algorithm, each party obtains its estimator gain by $K_i=A^{-1}\overline{H}^{-1}\overline{Z}_i$, and $K=\frac{1}{N}\begin{bmatrix}
K_1&K_2&\cdots&K_N
\end{bmatrix}$ is exactly the optimal steady-state Kalman gain $K^\star$ based on Theorem~\ref{thm:optimal}, which means that $\bar{x}(k)$ is the asymptotic MMSE estimate. At each iteration $t$, each party uploads its $\overline{Z}_i^t C_i$ and $\overline{Z}_i^t\sqrt{R_i}B_i$ to the cloud server and there is no need for party $i$ to publish its own observation parameters $C_i$ and $R_i$ directly. 

\begin{remark}
	In all of the four algorithms proposed in our paper, when $N\geq 3$, each party can hardly learn or infer any data or model information during the procedure, since each party receives the processed information from the cloud server and security module. The message broadcast to each party is a fusion of other parties' sensitive information. Other parties' privacy is preserved by the fusion processes. For example, in Algorithms~\ref{alg:secure} and~\ref{alg:gain}, the average is made public to all the $N$ parties. In Algorithms~\ref{alg:ADMM} and~\ref{alg:ADMM_2}, the message transmitted to party $i$ is dependent on other parties' parameters. To summarize, party $i$ can only infer the overall performance of the others. The sensitive information of each party is well preserved.   
\end{remark}

\section{Numerical Examples}\label{sec:numerical}
In this section, we illustrate the effective performance of our proposed secure multi-party state estimation with numerical examples. We consider a system with parameters:
\begin{align*}
A=\begin{bmatrix}
4.58&1.72&-0.54&-3.51&-0.14\\
2.77&2.07&-0.34&-2.68&-0.01\\
2.07&0.92&0.57&-2.15&0.19\\
5.36&2.46&-0.76&-4.20&-0.22\\
4.03&1.69&-0.29&-3.73&0.58
\end{bmatrix},~~Q=0.1I_5.
\end{align*}

\begin{figure}[t]
	\centering
	\includegraphics[width=0.49\textwidth]{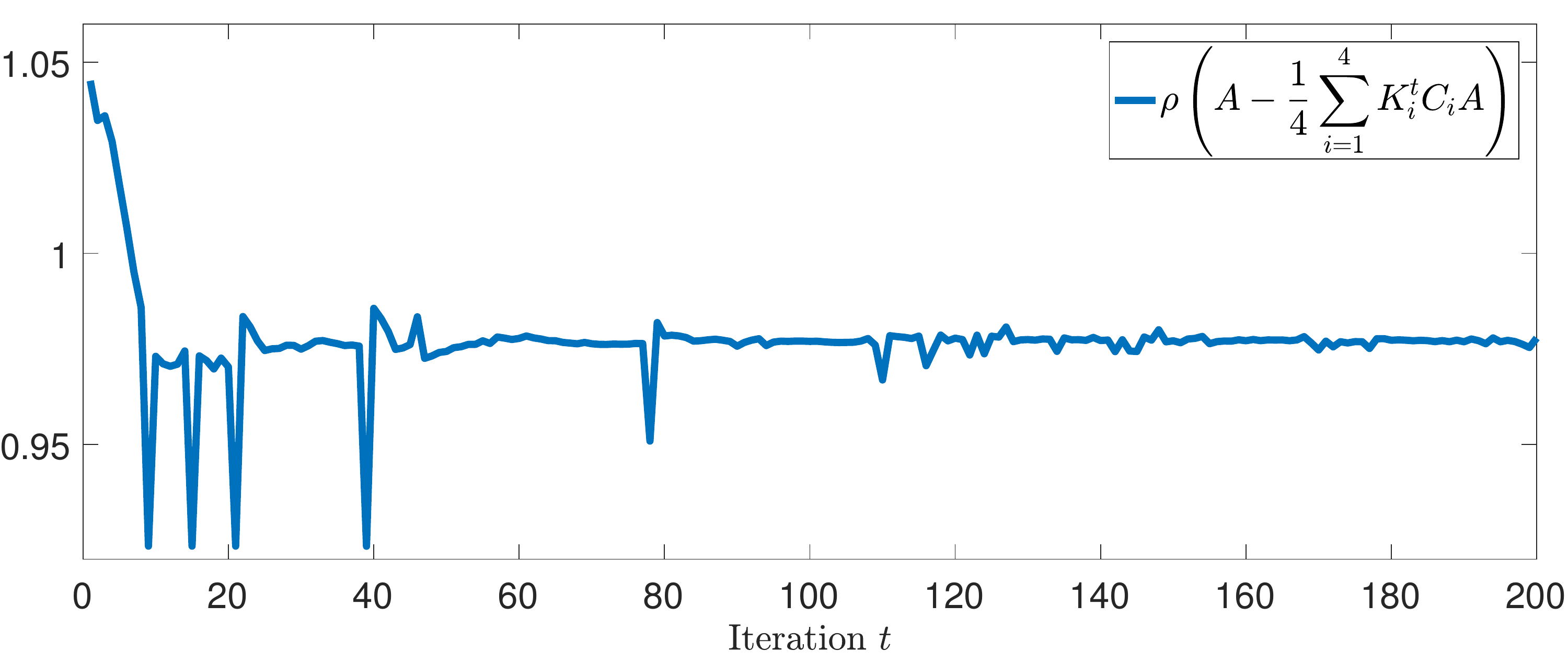}
	\caption{Stabilization design method II: ADMM for $N=4$.}
	\label{fig:ADMM}
\end{figure}

To illustrate the effectiveness of the stabilization design method II based on ADMM in Section~\ref{sec:ADMM}, we give an example where $N=4$. Four parties deploy their own sensor networks to monitor this process. The corresponding observation matrices are generated as follows:
\begin{align*}
C_1&=\begin{bmatrix}
0&0&1&0&0
\end{bmatrix},~~R_1=0.10,\\
C_2&=\begin{bmatrix}
0&1&0&0&0
\end{bmatrix},~~R_2=0.08,\\
C_3&=\begin{bmatrix}
1&0&0&1&0\\
1&1&0&0&0
\end{bmatrix},~~R_3=0.09I_2,\\
C_4&=\begin{bmatrix}
1&0&0&1&1
\end{bmatrix},~~R_4=0.06.
\end{align*}
In this scenario, even if $\displaystyle K_i=\argmin_{X\in\mathbb{R}^{n\times m_i}} \left\Vert \left(I_n-XC_i\right)A\right\Vert_2$, Inequality~\eqref{eqn:argmin} does not hold. We adopt Algorithm~\ref{alg:ADMM} to determine $K_i$'s. The parameter in the augmented Lagrangian is set as $\gamma=0.1$. The initial values $Z_i^0$ and $\Lambda_i^0$ are all set to $0$. The spectral radius of $\displaystyle A-\frac{1}{4}\sum_{i=1}^4 K_i^t C_i A$, which characterizes the stability of the proposed multi-party filtering protocol, is shown in Fig.~\ref{fig:ADMM}. After $200$ iterations, the obtained gains result in a spectral radius with value $0.98$.

\begin{figure}[t]
	\centering
	\includegraphics[width=0.49\textwidth]{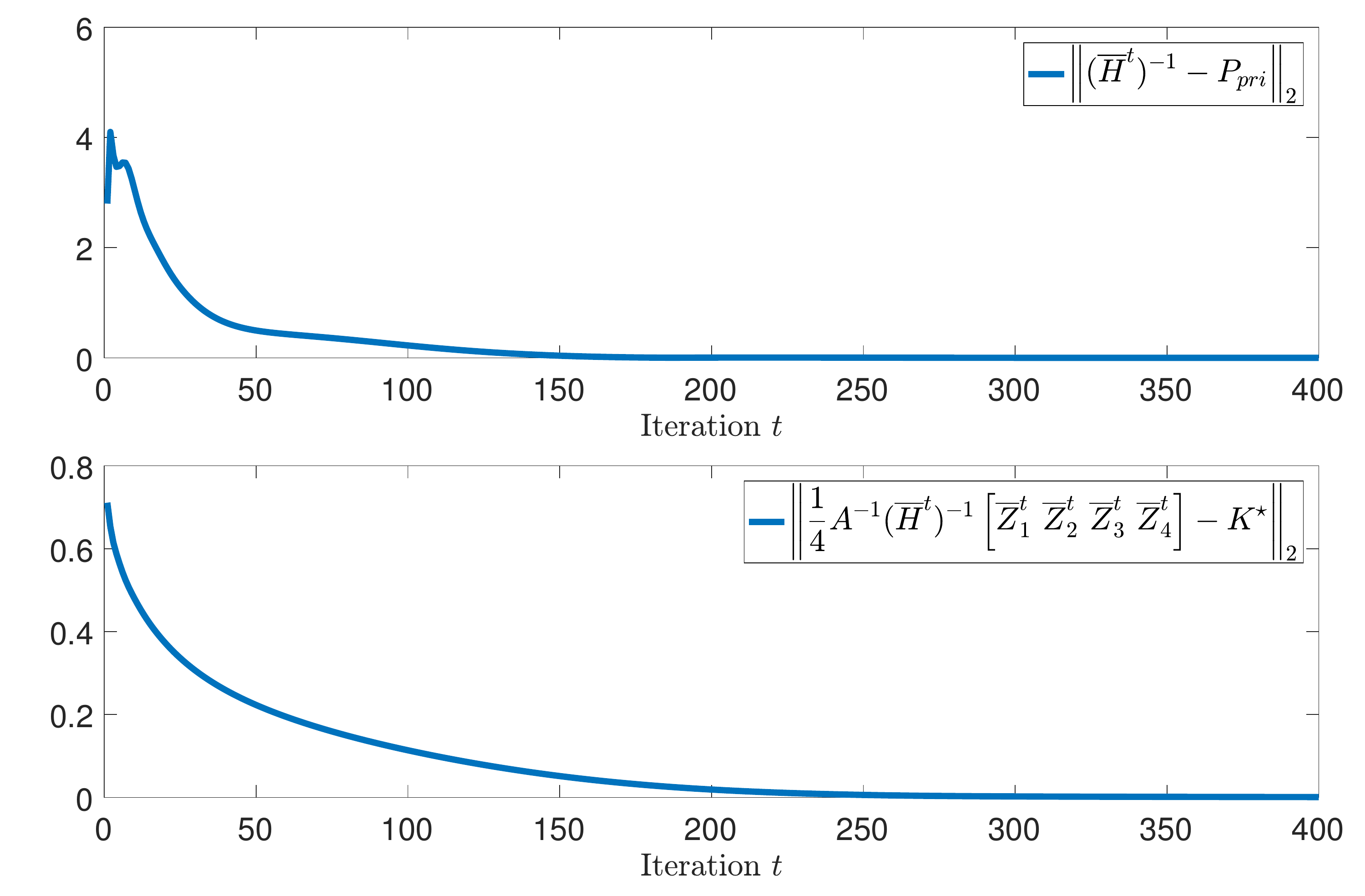}
	\caption{Asymptotic MMSE design method for $N=4$.}
	\label{fig:ADMM_2}
\end{figure}

To show the performance of Algorithm~\ref{alg:ADMM_2}, we use the same system and observation parameters. The related Lagrangian parameter $\gamma$ is set to be $0.3$, and all the initial values $\overline{Z}_i^0$ and $\overline{\Lambda}_i^0$ are set to be $0$. As portrayed in Fig.~\ref{fig:ADMM_2}, the estimator gain $\frac{1}{4} A^{-1}\left(\overline{H}^t \right)^{-1}\begin{bmatrix}
\overline{Z}_1^t&\overline{Z}_2^t&\overline{Z}_3^t&\overline{Z}_4^t
\end{bmatrix}$ calculated by Algorithm~\ref{alg:ADMM_2} approaches the optimal steady-state Kalman gain $K^\star$ after iterations, which verifies Theorem~\ref{thm:optimal}.

\begin{figure}[t]
	\centering
	\includegraphics[width=0.49\textwidth]{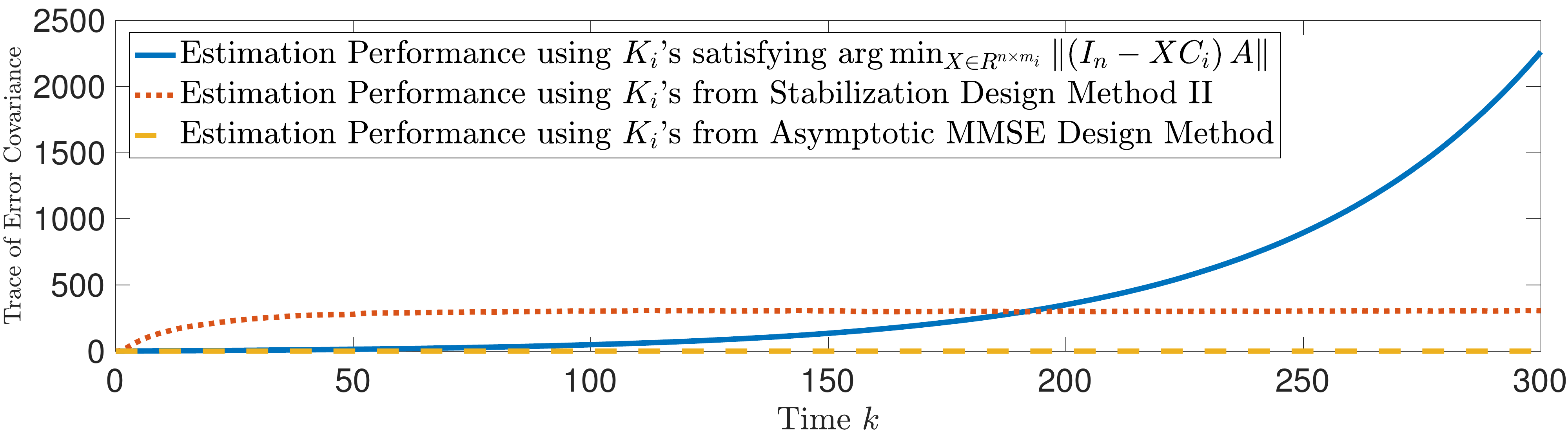}
	\caption{Trace of the state estimation error covariance using different $K_i$'s for $N=4$.}
	\label{fig:compare}
\end{figure}

To illustrate the difference caused by using different estimator gain $K_i$'s, we compare the trace of the estimation error covariance, i.e., $\Tr\{\overline{P}(k)\}$, when using $\displaystyle K_i=\argmin_{X\in\mathbb{R}^{n\times m_i}} \left\Vert \left(I_n-X C_i\right)A\right\Vert$, and $K_i$'s calculated by Algorithms~\ref{alg:ADMM} and~\ref{alg:ADMM_2}. Due to the existence of random noises, we run the simulation $20,000$ times for different $K_i$'s, respectively. As shown in Fig.~\ref{fig:compare}, $K_i$'s determined by the first method cannot stabilize the estimator, while $K_i$'s produced by the other two methods proposed in our paper can make the estimator stable. Furthermore, the yellow dashed line shows that the asymptotic MMSE design method achieves a good estimation performance, i.e., $\Tr\{\overline{P}(300)\}=0.27$.

\section{Conclusion and Future Work}\label{sec:con}
In this paper, we proposed a secure multi-party dynamic state estimation paradigm. During the whole procedure, the sensitive information of all the parties are well preserved. The local state estimates are protected by the additively \emph{homomorphic encryption}. Different methods were provided to collaboratively design the linear estimator gains with the purpose of guaranteeing the filtering stability without leakage of observation parameters.  Furthermore, an optimal gain design method was provided to reach the asymptotic MMSE estimation. Examples and simulations verified the theoretic results.

For future work, one possible direction is to provide privacy analysis on our proposed state estimation paradigm. For example, how to evaluate privacy breached by $Z_i C_i A$, $\overline{Z}_i C_i$, and $\overline{Z}_i \sqrt{R_i}B_i$ in Algorithms~\ref{alg:ADMM} and~\ref{alg:ADMM_2} would be worth exploring when the cloud server is assumed to be an adversary, or an honest but curious participant. One feasible method is to define a metric to characterize how accurate the estimates of $C_i$ and $R_i$ can be when the products are revealed. For security concerns, another possible direction is to investigate performance with imperfect communication channels and malicious adversaries. When delay and packet dropouts happen during the information transmission procedure, the estimation performance will be degraded inevitably. The alternative strategy of each party for uncertain channels is worth exploring. When there are malicious attackers or black sheep who deliberately tamper with the transmitted data to interfere with or even destroy the multi-party collaboration, it is necessary to design detection mechanisms to defend the collaboration and guarantee the performance. Furthermore, a more self-interested metric can be adopted as an extension. One may consider how to wisely fuse the local state estimate and the average state estimate for those parties who only care about their own estimation qualities in the privacy-preserving context.

\bibliographystyle{ieeetr}

\begin{IEEEbiography}
{Yuqing Ni} received the B.Eng. degree (Hons.) from the College of Control Science and Engineering, Zhejiang University, Hangzhou, China, and the Ph.D. degree in electronic and computer engineering from the Hong Kong University of Science and Technology, Hong Kong, in 2016, and 2020, respectively. From April 2019 to June 2019, she was a visiting student in the Department of Automatic, KTH Royal Institute of Technology, Stockholm, Sweden. Her research interests include security and privacy in cyber-physical system, networked state estimation, and wireless sensor networks.
\end{IEEEbiography}

\begin{IEEEbiography}
{Junfeng Wu} received the B.Eng. degree from the Department of Automatic Control, Zhejiang University, Hangzhou, China, and the Ph.D. degree in electrical and computer engineering from Hong Kong University of Science and Technology, Hong Kong, in 2009, and 2013, respectively. From September to December 2013, he was a Research Associate in the Department of Electronic and Computer Engineering, Hong Kong University of Science and Technology. From January 2014 to June 2017, he was a Postdoctoral Researcher in the ACCESS (Autonomic Complex Communication nEtworks, Signals and Systems) Linnaeus Center, School of Electrical Engineering, KTH Royal Institute of Technology, Stockholm, Sweden. He is currently with the College of Control Science and Engineering, Zhejiang University, Hangzhou, China. His research interests include networked control systems, state estimation, and wireless sensor networks, multi-agent systems. Dr. Wu received the Guan Zhao-Zhi Best Paper Award at the 34th Chinese Control Conference in 2015.	
\end{IEEEbiography}

\begin{IEEEbiography}
{Li Li} was born in April 1975, and is currently a professor and the chairman of Department of Control Science and Engineering, Tongji University. In 2003, she graduated from Shenyang Institute of Automation, Chinese Academy of Sciences with a doctorate degree in mechanical and electronic engineering. From February 2003 to January 2005, she was engaged in postdoctoral research in Tongji University. She is also Secretary-General of the Integrated Automation Technical Committee of the Chinese Automation Society, member of the Expert Consultation Committee of the Chinese Automation Society (ECC), member of the Shanghai Automation Society, member of the Shanghai Artificial Intelligence Society, and member of the Expert Committee of the Edge Computing Industry Alliance. The main research directions are data-driven modeling and optimization, computational intelligence, smart production, smart logistics and so on. She has presided over or participated in more than 10 projects of the National Key Research and Development Program of Science and Technology, National Natural Science Foundation of China, and Shanghai Science and Technology Innovation Action Plan. She published 3 academic monographs and more than 80 academic papers in high-level international and domestic journals and international conferences, and authorized 5 invention patents. She has won the first prize of Technical Invention Award of China Automation Society, the first prize of Shanghai Technical Invention Award and the first prize of Teaching Achievement of China Automation Society.
\end{IEEEbiography}

\begin{IEEEbiography}
{Ling Shi} received the B.S. degree in electrical and electronic engineering from Hong Kong University of Science and Technology, Kowloon, Hong Kong, in 2002 and the Ph.D. degree in Control and Dynamical Systems from California Institute of Technology, Pasadena, CA, USA, in 2008. He is currently a professor at the Department of Electronic and Computer Engineering, and the associate director of the Robotics Institute, both at the Hong Kong University of Science and Technology. His research interests include cyber-physical systems security, networked control systems, sensor scheduling, event-based state estimation, and exoskeleton robots. He is a senior member of IEEE. He served as an editorial board member for The European Control Conference 2013-2016. He was a subject editor for International Journal of Robust and Nonlinear Control (2015-2017). He has been serving as an associate editor for IEEE Transactions on Control of Network Systems from July 2016, and an associate editor for IEEE Control Systems Letters from Feb 2017. He also served as an associate editor for a special issue on Secure Control of Cyber Physical Systems in the IEEE Transactions on Control of Network Systems in 2015-2017. He served as the General Chair of the 23rd International Symposium on Mathematical Theory of Networks and Systems (MTNS 2018). He is a member of the Young Scientists Class 2020 of the World Economic Forum (WEF).
\end{IEEEbiography}

\end{document}